\documentclass[twocolumn]{cinc}
\usepackage{amsmath}
\usepackage{amssymb}
\usepackage{multirow}
\usepackage{colortbl}
\usepackage{comment}
\usepackage{balance}
\usepackage{setspace} 
\usepackage{subcaption}
\usepackage{hyperref}
\hypersetup{
    colorlinks = false,
    linkbordercolor = {1 0 0}}
\usepackage{url}

\usepackage{scalerel}
\usepackage{tikz}
\usetikzlibrary{svg.path}
\definecolor{orcidlogocol}{HTML}{A6CE39}
\tikzset{
  orcidlogo/.pic={
    \fill[orcidlogocol] svg{M256,128c0,70.7-57.3,128-128,128C57.3,256,0,198.7,0,128C0,57.3,57.3,0,128,0C198.7,0,256,57.3,256,128z};
    \fill[white] svg{M86.3,186.2H70.9V79.1h15.4v48.4V186.2z}
                 svg{M108.9,79.1h41.6c39.6,0,57,28.3,57,53.6c0,27.5-21.5,53.6-56.8,53.6h-41.8V79.1z M124.3,172.4h24.5c34.9,0,42.9-26.5,42.9-39.7c0-21.5-13.7-39.7-43.7-39.7h-23.7V172.4z}
                 svg{M88.7,56.8c0,5.5-4.5,10.1-10.1,10.1c-5.6,0-10.1-4.6-10.1-10.1c0-5.6,4.5-10.1,10.1-10.1C84.2,46.7,88.7,51.3,88.7,56.8z};
  }
}

\newcommand\orcidicon[1]{\href{https://orcid.org/#1}{\mbox{\scalerel*{
\begin{tikzpicture}[yscale=-1,transform shape]
\pic{orcidlogo};
\end{tikzpicture}
}{|}}}}

\begin{document}
\bibliographystyle{cinc}

\title{Early Myocardial Infarction Detection with One-Class Classification over Multi-view Echocardiography}

\author {Aysen Degerli$^{\dagger}$\orcidicon{0000-0002-9478-033X}, Fahad Sohrab$^{\dagger}$\orcidicon{0000-0002-8080-4011}, Serkan Kiranyaz$^{\ast}$\orcidicon{0000-0003-1551-3397}, Moncef Gabbouj$^{\dagger}$\orcidicon{0000-0002-9788-2323} \\
\ \\ % leave an empty line between authors and affiliation
$^\dagger$Tampere University, Tampere, Finland \\
$^\ast$Qatar University, Doha, Qatar
}

\maketitle

\begin{abstract}
Myocardial infarction (MI) is the leading cause of mortality and morbidity in the world. Early therapeutics of MI can ensure the prevention of further myocardial necrosis. Echocardiography is the fundamental imaging technique that can reveal the earliest sign of MI. However, the scarcity of echocardiographic datasets for the MI detection is the major issue for training data-driven classification algorithms. In this study, we propose a framework for early detection of MI over multi-view echocardiography that leverages one-class classification (OCC) techniques. The OCC techniques are used to train a model for detecting a specific target class using instances from that particular category only. We investigated the usage of uni-modal and multi-modal one-class classification techniques in the proposed framework using the HMC-QU dataset that includes apical $\textit{4-}$chamber (A$\textit{4}$C) and apical $\textit{2-}$chamber (A$\textit{2}$C) views in a total of $\textit{260}$ echocardiography recordings. Experimental results show that the multi-modal approach achieves a sensitivity level of $\textit{85.23\%}$ and F$\textit{1-}$Score of $\textit{80.21\%}$.
\end{abstract}

\section{Introduction}
\vspace{-0.1cm}
World Health Organization (WHO) has recently reported that coronary artery disease (CAD) is the reason for $16\%$ of total deaths worldwide \cite{factsheet2020top}. Myocardial infarction (MI) is the most severe manifestation of CAD that leads to irreversible necrosis of the myocardium \cite{reed2017acute}. Hence, early diagnosis of MI plays a vital role in the prevention of mortality and morbidity. Accordingly, the presentation of MI is recognized by its symptoms and several clinical features that are the biochemical markers, electrocardiography (ECG) findings, and imaging techniques \cite{thygesen2012third}. However, the symptoms of MI, i.e., shortness of breath and pain around the upper body, may not be visible in the early stages \cite{universal}. Furthermore, the biochemical values of myocardial necrosis, such as the high sensitivity cardiac troponin (hs-cTn), take time to evolve to a diagnostic level for MI \cite{macrae2006assessing, esmaeilzadeh2013role}. On the other hand, the changes at the ECG are occasionally non-diagnostic and also have a significant delay compared to imaging techniques \cite{esmaeilzadeh2013role}. Echocardiography is a non-invasive imaging technique that reveals the earliest sign of MI, which is the regional wall motion abnormality (RWMA) of the necrosed myocardium \cite{porter2018clinical}. Hence, echocardiography has the potential to be the most useful diagnostic tool to detect early MI with easy accessibility and low-cost~options \cite{chatzizisis2013echocardiographic}. 

The diagnosis of MI using echocardiography has several drawbacks, where the RWMA assessment is highly subjective, and the recordings generally have low image quality with a high level of noise \cite{porter2018clinical, 9354781}. Thus, computer-aided diagnosis algorithms have become a necessity for MI detection. However, many studies \cite{1407980, jamal2001noninvasive, 500138, omar2018automated} have evaluated their algorithms over scarce, private, synthetic, and single-view echocardiographic data which causes certain reliability and robustness issues, especially for deep learning models. Contrary to class-specific algorithms, one-class classification (OCC) models require only the positive class during training with much fewer samples \cite{9308359, sohrab2018subspace}. However, despite their feasibility, only the studies \cite{8861410, 10.1093/ehjci/ehz872.074} have used OCC models for echocardiographic data. 

\begin{figure*}[t!]
    \centering
    \includegraphics[width=\linewidth]{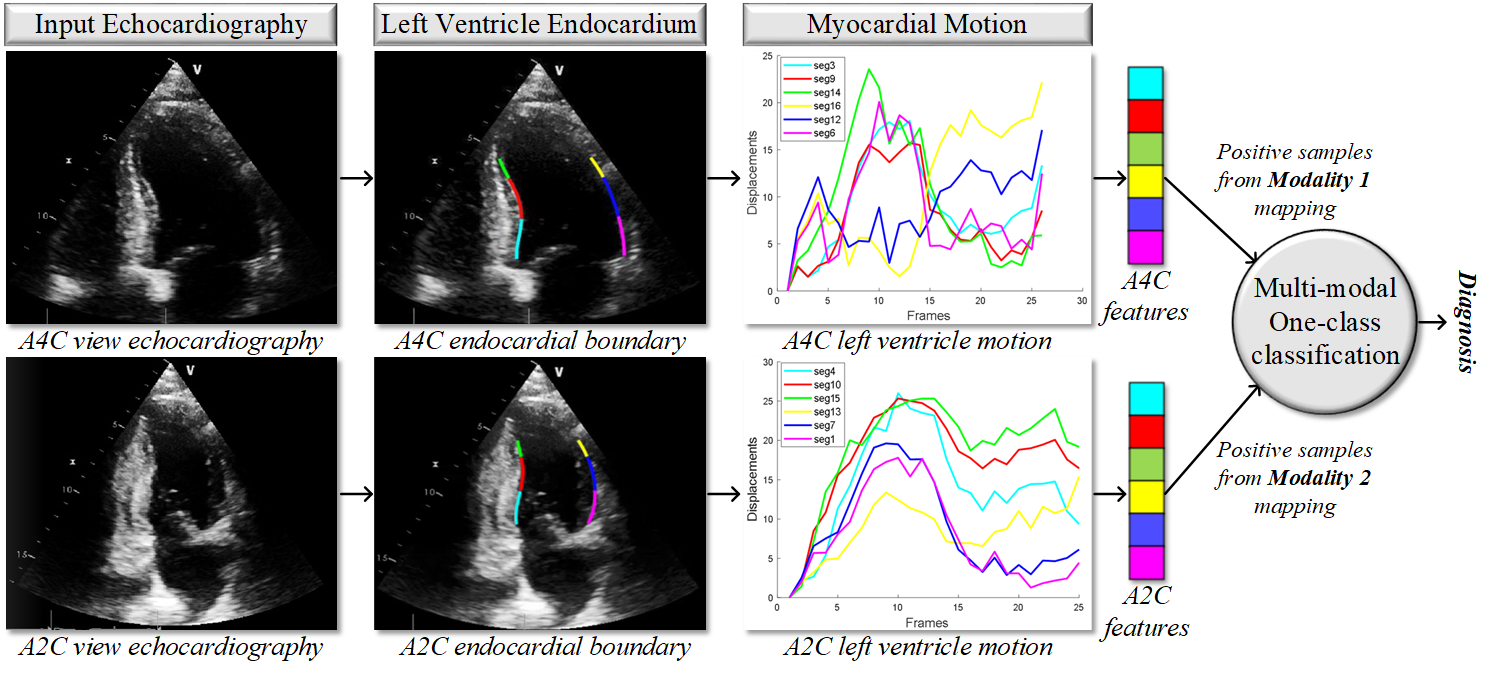}
    \caption{The proposed framework for early detection of MI with multi-modal one-class classification over multi-view echocardiography, where the features of both modalities are mapped to a shared subspace to perform diagnosis.}
    \label{fig:method}
\end{figure*}

In this study, we propose a framework that leverages OCC for the early detection of MI using multi-view echocardiography as depicted in Figure \ref{fig:method}. First, we extract features from apical 4-chamber (A4C), and apical 2-chamber (A2C) view echocardiography recordings by tracking the motion of the left ventricle (LV) using Active Polynomials (APs) \cite{degerli2021early}. Then, we use a multi-modal OCC approach over the maximum displacement features of A4C and A2C views. As the pioneer study with multi-modal OCC for the MI diagnosis using multi-view echocardiography, we have extensively evaluated both multi-, and uni-modal OCC algorithms over the HMC-QU\footnote{The benchmark HMC-QU dataset is publicly shared at the repository \href{https://www.kaggle.com/aysendegerli/hmcqu-dataset}{https://www.kaggle.com/aysendegerli/hmcqu-dataset}} dataset.

The paper proceeds as follows. In Section \ref{methodology}, we propose the framework for early MI detection. In Section \ref{experiments}, we report the experimental results and conclude the paper in Section \ref{conclusions}.

\section{Methodology}\label{methodology}
\vspace{-0.1cm}
The accurate extraction of the LV endocardium is a crucial step in myocardial motion tracking. In this study, the endocardial boundary of the LV from A4C and A2C views are extracted by Active Polynomials (APs) \cite{APs} that are the constrained versions of active contours \cite{kass1988snakes}. In order to overcome the common issues due to the low-quality in echocardiography, APs are formed by encapsulating the LV by a thick wall around the chamber, and then, the active contour is initialized and evolved towards the endocardium. Once the APs are formed over each frame of echocardiography recordings, the LV wall is divided into a total of $12$ distinct myocardial segments. Thus, myocardial motion is obtained for each myocardial segment as depicted in Figure \ref{fig:method}.

The feature engineering is performed as in our previous study \cite{degerli2021early}, where we extract the feature vectors from A4C and A2C view echocardiography recordings in one-cardiac cycle. In the proposed framework, we used OCC for training the predictive model. Contrary to class-specific algorithms, the OCC models do not require information from negative samples during training. For the OCC model, we propose using the Multi-modal Subspace Support Vector Data Description (MS-SVDD) \cite{SOHRAB2021107648} due to its feasibility with multi-view echocardiographic data. MS-SVDD maps the multi-view feature vectors to a lower-dimensional optimized feature space shared by features from different views of echocardiography as illustrated in Figure \ref{fig:method}. The feature vectors of view $v$ are represented by $\mathbf{F}$$_v$ $= [f_{v, 1}, f_{v, 2}, ..., f_{v,6}]$, $\mathbf{f}$$_{v, i}$ $\in \mathbb{R}$$^{D_v}$$\text{}$, where the dimensionality of the original feature space is $D_v$. Accordingly, a projection matrix $\mathbf{Q}$$_v$ $\in \mathbb{R}$$^{d \times D_v}$ is formed for each modality $v$ that projects the feature vectors $\mathbf{F}$$_v$$\text{}$ into a lower $d-$dimensional shared subspace optimized for OCC. Hence, MS-SVDD is trained by the target data that fits into the smallest hypersphere by minimizing the following function:
\begin{align}\label{minEq}
\min F(R,\textbf{a}) = R^2 + C\sum_{v=1}^{V}\sum_{i=1}^{N} \xi_{v,i} \nonumber \\ 
\text{s.t.   } \text{$||$$\mathbf{Q}$$_v$$\mathbf{f}$$_{v,i}$$ - \textbf{a}$$\|_2^2$$ \le R^2 + \xi _{v,i}, \xi_{v,i} \ge 0$},\nonumber\\ 
\forall v \in \{1,\dots,V\}, \forall i \in \{1,\dots,N\}, 
\end{align}

\noindent where $R$ is the radius, $\textbf{a}$ is the center of hypersphere, $\xi_{v,i}$ are slack variables, and $C$ controls the outliers in the training set. Then, $\mathbf{Q}$$_v$$\text{}$ is updated as $\mathbf{Q}\text{$_v$} \leftarrow \mathbf{Q}\text{$_v$} - \text{$\eta$} \text{$\Delta$} \text{$L$}$, where  $\Delta L$ is the gradient of Lagrangian of Eq. \eqref{minEq} for the corresponding modality $v$, and $\eta$ is the learning rate. Different regularization techniques ($r$) are also used in MS-SVDD by considering the co-variance of data from different modalities in the shared subspace. The regularization term expressing the co-variance of selected data is represented by $\omega$ and the importance of $\omega$ is controlled by a hyper-parameter $\beta$. 

\begin{table*}[t!]
\centering
\caption{Average myocardial infarction detection performance results (\%) computed over the test sets of each $5-$fold in HMC-QU dataset.}
\vspace{-0.4cm}
\resizebox{\textwidth}{!}{
\begin{tabular}{cccccccccccccccc}
& \multicolumn{7}{c}{Target: MI} &  & \multicolumn{7}{c}{Target: non-MI} \\
\cline{2-8} \cline{10-16}
& $r$ & Sen & Spe & Pre & F1 & Acc & GM &  & $r$ & Sen & Spe & Pre & F1 & Acc & GM  \\ 
\hline
\textit{Non-linear one-class classification} \\
\hline
$\text{MS-SVDD}_{ds_1}$ &$\omega_1$ & $70.45$ & $61.90$ & $\bold{79.49}$ & $74.70$ & $67.69$ & $\bold{66.04}$ & & $\omega_0$& $71.43$ & $53.41$ & $42.25$ & $53.10$ & $59.23$ & $61.77$ \\

$\text{MS-SVDD}_{ds_2}$ &$\omega_2$ & \cellcolor[gray]{.96}$63.64$ & \cellcolor[gray]{.96}$42.86$ & \cellcolor[gray]{.96}$70.00$ & \cellcolor[gray]{.96}$66.67$ & \cellcolor[gray]{.96}$56.92$ & \cellcolor[gray]{.96}$52.23$ & &$\omega_0$ & \cellcolor[gray]{.96}$61.90$ & \cellcolor[gray]{.96}$73.86$ & \cellcolor[gray]{.96}$53.06$ & \cellcolor[gray]{.96}$57.14$ & \cellcolor[gray]{.96}$70.00$ & \cellcolor[gray]{.96}$67.62$ \\

$\text{MS-SVDD}_{ds_3}$ &$\omega_2$ & $55.68$ & $59.52$ & $74.24$ & $63.64$ & $56.92$ & $57.57$ & &$\omega_0$ & $57.14$ & $57.95$ & $39.34$ & $46.60$ & $57.69$ & $57.54$\\

$\text{MS-SVDD}_{ds_4}$ &$\omega_6$ & \cellcolor[gray]{.96}$39.77$ & \cellcolor[gray]{.96}$\bold{76.19}$ & \cellcolor[gray]{.96}$77.78$ & \cellcolor[gray]{.96}$52.63$ & \cellcolor[gray]{.96}$51.54$ & \cellcolor[gray]{.96}$55.05$ & &$\omega_2$ & \cellcolor[gray]{.96}$73.81$ & \cellcolor[gray]{.96}$67.05$ & \cellcolor[gray]{.96}$51.67$ & \cellcolor[gray]{.96}$\bold{60.78}$ & \cellcolor[gray]{.96}$69.23$ & \cellcolor[gray]{.96}$\bold{70.35}$ \\

$\text{ES-SVDD}$ &$\psi_0$ & $73.86$ & $38.10$ & $71.43$ & $72.63$ & $62.31$ & $53.05$ & &$\psi_0$ & $69.05$ & $56.82$ & $43.28$ & $53.21$ & $60.77$ & $62.64$ \\

$\text{S-SVDD}$ &$\psi_2$ & \cellcolor[gray]{.96}$59.09$ & \cellcolor[gray]{.96}$54.76$ & \cellcolor[gray]{.96}$73.24$ & \cellcolor[gray]{.96}$65.41$ & \cellcolor[gray]{.96}$57.69$ & \cellcolor[gray]{.96}$56.88$ & &$\psi_1$ & \cellcolor[gray]{.96}$54.76$ & \cellcolor[gray]{.96}$52.27$ & \cellcolor[gray]{.96}$35.38$ & \cellcolor[gray]{.96}$42.99$ & \cellcolor[gray]{.96}$53.08$ & \cellcolor[gray]{.96}$53.50$ \\

SVDD &$-$& $80.68$ & $38.10$ & $73.20$ & $76.76$ & $66.92$ & $55.44$ & & $-$& $69.05$ & $71.59$ & $53.70$ & $60.42$ & $\bold{70.77}$ & $70.31$ \\

OC-SVM &$-$& $42.05$ & $71.43$ & $75.51$ & $54.01$ & $51.54$ & $54.81$ & &$-$ & $35.71$ & $\bold{82.95}$ & $50.00$ & $41.67$ & $67.69$ & $54.43$ \\

\hline 
\textit{Linear one-class classification}\\
\hline
$\text{MS-SVDD}_{ds_1}$ &$\omega_5$ & $81.82$ & $47.62$ & $76.60$ & $79.12$ & $70.77$ & $62.42$ &  & $\omega_2$ & $73.81$ & $62.50$ & $48.44$ & $58.49$ & $66.15$ & $67.92$\\

$\text{MS-SVDD}_{ds_2}$ & $\omega_2$& \cellcolor[gray]{.96}$54.55$ & \cellcolor[gray]{.96}$59.52$ & \cellcolor[gray]{.96}\cellcolor[gray]{.96}$73.85$ & \cellcolor[gray]{.96}$62.75$ & \cellcolor[gray]{.96}$56.15$ & \cellcolor[gray]{.96}$56.98$ & &$\omega_2$ & \cellcolor[gray]{.96}$78.57$ & \cellcolor[gray]{.96}$36.36$ & \cellcolor[gray]{.96}$37.08$ & \cellcolor[gray]{.96}$50.38$ & \cellcolor[gray]{.96}$50.00$ & \cellcolor[gray]{.96}$53.45$ \\

$\text{MS-SVDD}_{ds_3}$ &$\omega_0$ & $67.05$ & $59.52$ & $77.63$ & $71.95$ & $64.62$ & $63.17$ & &$\omega_5$ & $73.81$ & $50.00$ & $41.33$ & $52.99$ & $57.69$ & $60.75$\\

$\text{MS-SVDD}_{ds_4}$ &$\omega_5$ & \cellcolor[gray]{.96}$85.23$ & \cellcolor[gray]{.96}$42.86$ &\cellcolor[gray]{.96}$75.76$ & \cellcolor[gray]{.96}$\bold{80.21}$ & \cellcolor[gray]{.96}$\bold{71.54}$ & \cellcolor[gray]{.96}$60.44$ & & $\omega_0$& \cellcolor[gray]{.96}$\bold{80.95}$ & \cellcolor[gray]{.96}$59.09$ & \cellcolor[gray]{.96}$48.57$ & \cellcolor[gray]{.96}$60.71$ & \cellcolor[gray]{.96}$66.15$ & \cellcolor[gray]{.96}$69.16$ \\

$\text{ES-SVDD}$ &$\psi_3$ & $82.95$ & $35.71$ & $73.00$ & $77.66$ & $67.69$ & $54.43$ & &$\psi_3$ & $45.24$ & $67.05$ & $39.58$ & $42.22$ & $60.00$ & $55.08$ \\

$\text{S-SVDD}$ & $\psi_3$& \cellcolor[gray]{.96}$70.45$ & \cellcolor[gray]{.96}$45.24$ & \cellcolor[gray]{.96}$72.94$ & \cellcolor[gray]{.96}$71.68$ & \cellcolor[gray]{.96}$62.31$ & \cellcolor[gray]{.96}$56.45$ & &$\psi_2$ & \cellcolor[gray]{.96}$50.00$ & \cellcolor[gray]{.96}$70.45$ & \cellcolor[gray]{.96}$44.68$ & \cellcolor[gray]{.96}$47.19$ & \cellcolor[gray]{.96}$63.85$ & \cellcolor[gray]{.96}$59.35$ \\

SVDD & $-$& $\bold{86.36}$ & $33.33$ & $73.08$ & $79.17$ & $69.23$ & $53.65$ & & $-$ & $69.05$ & $69.32$ & $51.79$ & $59.18$ & $69.23$ & $69.18$ \\

OC-SVM &$-$ & \cellcolor[gray]{.96}$44.32$ & \cellcolor[gray]{.96}$73.81$ & \cellcolor[gray]{.96}$78.00$ & \cellcolor[gray]{.96}$56.52$ & \cellcolor[gray]{.96}$53.85$ & \cellcolor[gray]{.96}$57.19$ & & $-$ & \cellcolor[gray]{.96}$47.62$ & \cellcolor[gray]{.96}$81.82$ & \cellcolor[gray]{.96}$\bold{55.56}$ & \cellcolor[gray]{.96}$51.28$ & \cellcolor[gray]{.96}$\bold{70.77}$ & \cellcolor[gray]{.96}$62.42$ \\

\end{tabular}}
\label{tab:results}
\end{table*} 

In this study, we also investigate the uni-modal OCC algorithms: One-class Support Vector Machine (OC-SVM) \cite{scholkopfu1999sv}, Support Vector Data Description (SVDD) \cite{tax2004support}, Subspace SVDD (S-SVDD) \cite{sohrab2018subspace}, and Ellipsoidal Subspace SVDD (ES-SVDD) \cite{9133428}. Contrary to MS-SVDD, where the feature vectors are projected to a joint subspace suitable for OCC, in the uni-modal OCC methods, we concatenate the feature vectors of A4C and A2C
views as $\mathbf{F} = \begin{bmatrix} $$\mathbf{F}$$_1$$\text{ }$$\mathbf{F}$$_2$$ \end{bmatrix} \in $ $\mathbb{R}$$^{(D_1+D_2) \times N}$. In uni-modal subspace OCC methods (S-SVDD, ES-SVDD), the corresponding regularization technique is denoted by~$\psi$.

\section{Experimental Evaluation}\label{experiments}
\vspace{-0.1cm}
In this section, the experimental setup is introduced. Then, the experimental results are reported over the HMC-QU dataset.

\subsection{Experimental Setup}
\vspace{-0.1cm}
The performance of the proposed framework is evaluated over the HMC-QU dataset \cite{degerli2021early} that includes a total of 260 echocardiography recordings from A4C and A2C views of 130 individuals with the ground-truths of $88$ MI patients, and $42$ non-MI subjects. During the training of the OCC models, we consider the target class as MI or non-MI, and report the results for both targets. Accordingly, noting that the target class is the positive class, we calculate the standard performance metrics as follows: Sensitivity (Sen) is the ratio of correctly detected positive samples in the positive class, Specificity (Spe) is the rate of accurately identified negative samples in the negative class, Precision (Pre) is the ratio of correctly detected target samples among the samples that are identified as the positive class, F$1-$Score (F1) is the harmonic mean of \textit{Sen} and \textit{Pre}, Accuracy (Acc) is the ratio of correctly classified samples over the dataset, and GMean (GM) is the geometric mean of \textit{Sen} and \textit{Spe}. 

The OCC models are evaluated in a stratified $5-$fold cross-validation (CV) scheme with a ratio of $80\%$ training to $20\%$ test sets. The best hyper-parameters for the testing phase are determined by an exhaustive search over a stratified $10-$fold CV scheme with respect to the best GM during training. We have experimented with both linear and non-linear (kernel) versions of the OCC models, where we used the kernel $K_{i,j}=\exp \left(\frac{-||\mathbf{f}\text{$_i$}-\mathbf{f}\text{$_j$}||^\text{$2$}}{2\sigma ^2}\right)$ with the hyper-parameter $\sigma$. The hyper-parameters $\eta$, $\beta$, $C$, $\sigma$, and $d$ are searched as follows: $\eta \in \{10^{-4}, 10^{-3}, 10^{-2}, 10^{-1}, 1\}$, $\beta \in \{10^{-4}, 10^{-3}, 10^{-2}, 10^{-1}, 1, 10, 10^{2}, 10^{3}, 10^{4}\}$, $C \in \{0.01, 0.05, 0.1, 0.2, 0.3, 0.4, 0.5, 0.6\}$, $\sigma \in \{10^{-2}, 10^{-1}, \\ 1, 10, 10^{2}, 10^{3}\}$, in multi-modal $d \in [1, 5]$, whereas in uni-modal $d \in [1, 11]$ with a gap of $1$ increasing at each step. Moreover, the MS-SVDD has different decision strategies $ds_1, ds_2, ds_3, ds_4$, where the details are presented in \cite{SOHRAB2021107648}. Lastly, the implementation of the OCC models is performed on MATLAB R2020a.

\begin{table}[b!]
\centering
\caption{Confusion matrices of the linear SVDD (a) and $\text{MS-SVDD}_{ds_4}$ (b) models with target MI.}
\vspace{-0.2cm}
\begin{subtable}{.48\linewidth}
\centering
\caption{}
\vspace{-0.2cm}
\resizebox{\linewidth}{!}{
\begin{tabular}{|c|c|c|c|}
\hline
\multicolumn{2}{|c|}{\multirow{2}{*}{\textbf{SVDD}}} & \multicolumn{2}{c|}{Predicted} \\ \cline{3-4} 
\multicolumn{2}{|c|}{} & \multicolumn{1}{c|}{Non-MI} & \multicolumn{1}{c|}{MI} \\ \hline
\multirow{2}{*}{\begin{tabular}[c]{@{}c@{}}Ground\\ Truth\end{tabular}} & Non-MI & $14$ & $28$ \\ \cline{2-4} 
 & MI & $12$ & $76$ \\ \hline
\end{tabular}}
\end{subtable}
\vspace{-0.2cm}
\bigskip
\noindent
\begin{subtable}{.48\linewidth}
\centering
\caption{}
\vspace{-0.2cm}
\resizebox{\linewidth}{!}{
\begin{tabular}{|c|c|c|c|}
\hline
\multicolumn{2}{|c|}{\multirow{2}{*}{\textbf{$\text{MS-SVDD}_{ds_4}$}}} & \multicolumn{2}{c|}{Predicted} \\ \cline{3-4} 
\multicolumn{2}{|c|}{} & \multicolumn{1}{c|}{Non-MI} & \multicolumn{1}{c|}{MI} \\ \hline
\multirow{2}{*}{\begin{tabular}[c]{@{}c@{}}Ground\\ Truth\end{tabular}} & Non-MI & $18$ & $24$ \\ \cline{2-4} 
 & MI & $13$ & $75$ \\ \hline
\end{tabular}}
\end{subtable}
\label{tab:CMs}
\end{table}

\subsection{Experimental Results}
\vspace{-0.1cm}
In this section, we investigate the performances of multi- and uni-modal OCC models for different targets with linear and non-linear versions. The performances are reported in Table \ref{tab:results}. Primarily, the best GM of $66.04\%$ and $70.35\%$ are obtained by non-linear MS-SVDD for MI and non-MI targets, respectively. It can be observed that non-linear $\text{MS-SVDD}_{ds_1}$ has achieved the highest precision of $79.49\%$ for target MI, where the decision strategy $1$ is performed that merges the decisions of both modalities by the \textit{AND} operator in the testing phase. Moreover, the best F$1-$Score of $80.21\%$ is achieved by linear $\text{MS-SVDD}_{ds_4}$ for target MI with an elegant sensitivity level of $85.23\%$, where only the decision of the second modality is considered in the testing phase. The best sensitivity level of $86.36\%$ is obtained by linear SVDD for target MI which is very close to the sensitivity of linear $\text{MS-SVDD}_{ds_4}$, where their confusion matrices are shown in Table \ref{tab:CMs}. 

\section{Conclusions}\label{conclusions}
\vspace{-0.1cm}
The early diagnosis of MI is a crucial task to prevent the further myocardial necrosis. This study investigates the OCC algorithms for the first time for multi-view echocardiography. The experimental results over the HMC-QU dataset have revealed that multi-modal OCC models have achieved the highest precision of $79.49\%$ and F$1-$Score of $80.21\%$ despite the decent performance of the uni-modal OCC algorithms. Furthermore, we have investigated the linear and non-linear options of the presented OCC algorithms, and experimentally showed that the best GMean of $70.35\%$ is achieved by the multi-modal OCC model. 

\section*{Acknowledgments}
\vspace{-0.1cm}
This study was supported in part by the NSF-Business Finland Center for Visual and Decision Informatics (CVDI) Advanced Machine Learning for Industrial Applications (AMaLIA) under Grant 4183/31/2021, and in part by the Haltian Stroke-Data projects.

\begin{spacing}{0.86}
\bibliography{refs}
\end{spacing}

\begin{correspondence}
Aysen Degerli\\
P.O. Box $553$, FI$-33014$, Tampere Finland\\
aysen.degerli@tuni.fi
\end{correspondence}

%\balance

\end{document}